\documentclass[useAMS,usenatbib]{mn2e}
\usepackage{graphicx,times}
\usepackage{amssymb}
\usepackage{amsmath}
\usepackage{longtable}
\usepackage{rotating}
\title[Binary interactions on parameter determinations]{The effects of binary interactions on parameter determinations for early-type galaxies}
\author[Zhang et al.]
{Yu~Zhang$^{1}$\thanks{E-mail: zhy@xao.ac.cn},
Jinzhong~Liu$^{1}$, Fenghui~Zhang$^{2}$
\\
$^{1}$Xinjiang Astronomical Observatory, Chinese Academy of Sciences, Urumqi 830011, China\\
$^{2}$Yunnan Observatory, Chinese Academy of Sciences, Kunming 650011, China
}

\begin{document}


\maketitle

\label{firstpage}

\begin{abstract}
Based on stellar population models without (SSP) and with (BSP) binary interactions, we investigate the effects of binary interactions on parameter determinations for early-type galaxies (ETGs).
We present photometric redshift (photo-$z$), age and spectral type for photometric data sample by fitting observed magnitudes with the SSP and BSP models. Our results show that binary interactions have no effect on photo-$z$ estimation. Once we neglect binary interactions, the age of ETGs will be  underestimated, by contrast, the effects on the age estimations can be negligible for other type of galaxies.
For ETG sample, we derive their properties by fitting their spectra with the SSP and BSP models. When comparing these galaxy properties, we find no variation of the overall metallicities for ETGs among the SSP and BSP models. Moreover, the inclusion of binary interactions can affect age estimations. Our results show that the BSP-fitted ages in $\sim33.3\%$ of ETG sample are around $0.5-1.0$\,Gyr larger than the SSP-fitted ages; $\sim44.2\%$ are only $0.1-0.5$\,Gyr larger; the rest $\sim22.5\%$ are approximately equal.
By comparisons, we find the difference of the star formation rate between the SSP and BSP models is large at the late evolution stage.

\end{abstract}


\section{INTRODUCTION}
\label{sect:intro}
Early-type galaxies (ETGs), which comprise elliptical and lenticular (S0) galaxies, are the oldest class of galaxy. 
And their stellar populations formed at early time \citep{Trager00,Temi05,DeLucia06} and passively evolved to their present styles. 
The phenomenon of far-ultraviolet (FUV) excess in ETGs became  surprising since its first discovery by the $Orbitting\,\, Astronomical\,\, Observatory-2$ in $1969$ \citep{code72,burstein88}. The flux in the spectral energy distributions (SEDs) of ETGs increases with a decreasing wavelength in the range from $2000$ to $1200\rm \AA$. This behavior is known as UV upturn, UV rising-branch, or UVX \citep[see][for a review]{oconnell99}. And some recent studies show evidence that ETGs have some current events of minor star formation \citep{Kaviraj07,Schawinski07,Salim12,Barway13}, which contribute the FUV spectra.

The favored origin of the UV-upturn is extreme horizontal branch (EHB) stars and their descendants, either metal poor \citep{Lee94,Park97} or metal rich \citep{Bressan94,Yi97}. And the detection of EHB stars in the dwarf elliptical galaxy M32 provide direct evidence for the EHB origin of the UV-upturn.
Meanwhile, binary evolution can also reproduce EHB stars \citep{Han02,Han03}. \citet{Han07} have used three possible binary evolution channels for EHB stars to explain the UV-upturn in ETGs, with limited dependence on age and metallicity. \citet{Hernandez14} also used the stellar population model with binary stars to study the UV-upturn of ETGs, and found that the UV-upturn was very sensitive to the fraction of binary stars.

Binary stars are very common in galaxies, and the evolution of binary stars is different from that of single stars. Meanwhile, observations also show that binary stars are common in nearby star clusters and galaxies \citep{abt83,Carney05,Sollima07,Raghavan10,Minor13}.  \citet{Brinchmann10} also indicated that the importance of binary stars was one of six important challenges in stellar population studies in the next decades. However, only few works had done for studying the influence of binary evolution on stellar population properties. \citet{Zhang04,Zhang05} have shown that the UV passbands could be about $2.0-3.0$ mag enhanced once binary interactions have been taken into account. Therefore, ignoring the binary interactions in evolution population synthesis (EPS) models can underestimate the UV flux and affect the property determinations for stellar population systems \citep{zfh12,Zhang12,Zhang13}

While the importance of binary evolution has been shown, detailed studies of the effects of binary stars on property determinations of ETGs are not enough. Therefore, we investigate the effects of binary interactions on parameter determinations for ETGs based on the EPS models with and without binary interactions.
 At the first step, we use a standard SED fitting procedure to fit the SED of nearby galaxy photometric data sample based on the EPS models with and without binary interactions and determine their photometric redshifts (photo-$z$), ages and spectral types.
We also study the effects of binary interactions on the estimations of photometric redshift and age. 
In order to study the effects of binary interactions on the property determinations of ETGs, we use the full spectrum-fitting technique based on the EPS models with and without binary interactions to fitting the spectra of ETGs sample. Our main aim is to investigate the effects of binary interactions on the property determinations for ETGs.

The paper is organized as follows. In Section $2$, we describe the galaxy sample used in this study. The method and models are given in Section $3$. The analysis of the effects of binary interactions follow in Sections $4$ and $5$. We show the summary and conclusion in Section $6$.

\section{GALAXY SAMPLE}
At first, in order to investigate the effects of binary interactions on phto-$z$ determined by using SED fitting method, one needs to construct the galaxy sample with known spectroscopic redshifts and multi-band magnitudes. We select galaxies in the photometric data sample from the Sloan Digital Sky Survey (SDSS) Data Release 7 (DR7), which include the spectroscopic redshifts and optical magnitudes ($ugriz$). Considering the computational time, $11000$ galaxies have been randomly selected. Then, we correlate the $Galaxy Evolution Explorer$ ($GALEX$) Data Release 4 (DR4) source positions with positions from the SDSS DR7 catalogue.  The matching radius between SDSS DR7 and $GALEX$ DR4 is $6^{''}$.  Finally, the final photometric data sample catalogue includes $4278$ galaxies with all $F_{UV}$, $N_{UV}$ and $ugriz$ magnitudes.

We also investigate the effect of binary interactions on the stellar population properties of ETGs. The ETG sample is selected from \citet{Fukugita07}, which provides a catalogue of morphological classified galaxies and consists $2253$ galaxies with $r$ band Petrosian magnitude brighter than $16$ mag. Among these sample, $323$ galaxies belong to ETGs (with $T= 0$). Removing the objects mismatched with SDSS DR7 and $GALEX$ DR4, $120$ ETGs with both photometrical ($F_{UV}$, $N_{UV}$ and $ugriz$ magnitudes) and spectroscopic information are left. For these $120$ galaxies, the $F_{UV}$ and $N_{UV}$ magnitudes are obtained from $GALEX$ DR4, and the $ugriz$ magnitudes and spectra are selected from SDSS DR7. 

\section{THE METHODS AND MODELS}

\subsection{The methods}
\subsubsection{SED fitting method}
In our work, the photo-$z$ is obtained by the HyperZ code of \citet{Bolzonella00}, which is the first publicly available phto-$z$ code and has been widely used for photo-$z$ estimations of galaxies \citep{Maraston10,Abdalla11}. HyperZ uses a template-fitting algorithm and adopts a standard $\chi^{2}$ minimization procedure:
\begin{equation}
\label{eq:hyperz_1}
\chi^{2}=
\sum\limits_{i=1}^{N_{\rm{filter}}}\left[ {{F\rm_{\rm{obs,}\emph{i}}- \emph{bF}_{\rm{temp,}\emph{i}}(\emph{z})}\over{\sigma_{i}}} \right]^2,
\end{equation}
where $F_{\rm{obs,}\emph{i}}$, $F_{\rm{tmp,}\emph{i}}$, and $\sigma_{i}$ are the observed and template fluxes and their uncertainty in filter $i$, respectively, and $b$ is a normalization constant. 

HyperZ is based on the standard SED fitting technique and the detection of strong spectral features, such as the $4000 \rm\AA$ break, the Balmer break, the Lyman decrement or strong emission lines.

The inputs for HyperZ are the filter set and the photometric catalogue of galaxies, which includes magnitudes and photometric errors through the filters specified in the filter set. For a given galaxy catalogue, several relevant parameters in the photo-$z$ calculation procedure are included.

(i) The set of template spectra, which includes the type of star formation rate (SFR), the possible link between the age and metallicity of the stellar populations, and the choice of an initial mass function (IMF). The template spectral library will be presented in Section $3.2.2$.

(ii) HyperZ presents five forms of reddening law, and the reddening raw of \citet{Calzetti00} is used in this work. The $A_{V}$ is allowed to vary from $0.0$ to $1.2$ in steps of 0.2, which corresponding to $E(B-V)$ varying from $0.0$ to $0.30$ according to the reddening law ($R\rm_{V} = 4.05 $) of \citet{Calzetti00}.

(iii) The flux decrements in Lyman forest are computed according to \citet{Giallongo90} and \citet{Madau95}.

(iv) The filter involved in this study include the $GALEX$ and the SDSS standard filter system $F_{UV}$, $N_{UV}$, and $ugriz$. And the limiting magnitudes in each filters are set to $29.0, 29.0, 29.0, 30.0, 30.0, 30.0$ and $30.0$ mag, respectively.

(v) We adopt a set of standard dark matter model cosmology parameters ($\Omega\rm_{\Lambda}, \Omega\rm_{M}, H_{0}$) $= 0.7, 0.3, 70.0$.

The outputs of Hyperz are photo-$z$, age and spectral type.
The descriptions of photometric data of galaxies are given in Section $2$. The template spectral library is presented in Section $3.2.2$.

\subsubsection{The spectrum-fitting method}
 
In order to obtain the age, metallicity and star formation history (SFH) of ETGs, we apply full spectrum-fitting method of the STARLIGHT code by \citet{CidFernandes05}. 
The STARLIGHT code is originally used to study the properties
of galaxy, and is achieved by fitting the observed spectrum \emph{F$\rm_{O}$} with a model spectrum
\emph{F$\rm_{M}$} that mixed by \emph{N$_{\star}$} SSPs with
different ages and metallicities from the EPS models. The code is carried out with a simulated
annealing plus the Metropolis scheme \citep{CidFernandes01}, which
searches for the minimum
\begin{equation}
\label{eq:starlight_1}
\;\;\;\;\;\;\;\;\;\;\;\;\;\chi^2= \sum\limits_{\lambda}\,
[(F\rm_{O}-\emph{F}\rm_{M})\omega_{\lambda}]^2,
\end{equation}
where $\omega^{-1}_{\lambda}$ is the error on
\emph{F$\rm_{O}$}. The
line-of-sight stellar motions are modeled by a Gaussian distribution
centered at velocity \emph{v}$_{\star}$ and with dispersion
\emph{$\sigma$}$_{\star}$. Furthermore, in this process, we construct a base with \emph{N$_{\star}=196$} stellar populations, encompassing $28$ ages between $0.001$ and $12.6$ Gyr and 7 metallicities: $Z= 0.0001, 0.0003, 0.001, 0.004, 0.01, 0.02$, and $0.03$. These stellar population models are obtained from the EPS models, which are given in Section 3.2.1.

The outputs includes several physical properties, such as the stellar mass, stellar extinction, mean stellar age, mean metallicity as well as full time-dependent star formation. Following \citet{CidFernandes05}, we obtain the light- and mass-weighted mean stellar ages defined as 
\begin{equation}
\label{eq:starlight_2}
\langle {\rm log}t_{\star} \rangle_{L}= \sum\limits_{j=1}^{N_{\star}}
x_{j}{\rm log}t_{j},
\end{equation}
and
\begin{equation}
\label{eq:starlight_3}
\langle {\rm log}t_{\star} \rangle_{M}= \sum\limits_{j=1}^{N_{\star}}
\mu_{j}{\rm log}t_{j},
\end{equation}
where $x_{j}$ and $\mu_{j}$ are the $population \,\, vector$ (the fraction of light contributed by certain SSP) and the mass fraction vector (the fraction of stellar mass contributed by each SSP), respectively. The light- and mass-weighted average metallicity are obtained in the same way.

The SFHs can provide insight into mechanisms governing the formation and evolution of galaxies. To obtain the SFHs of the ETGs sample, we use the STARLIGHT code  with EPS models to fit the observed spectra and select the best population mixture. The SFR as a function of time can be used to characterize the SFH of galaxy.
The STARLIGHT code dose not provide immediately the SFR after fitting, but it gives the optimal weights of the different stellar population components. Just as the approach described by \citet{Asari07} and \citet{Koleva09}, we use the smoothed strategy to derive the smoothed SFR. 

Following \citet{Asari07}, we can obtain the smoothed SFR as,

\begin{equation}
\label{eq:SFR}
       {\rm SFR}(t_\star)  \approx
       \frac{\Delta M^c_\star(t_\star)}{\Delta t_\star} =
       \frac{M_\star^c \log e}{t_\star}\frac {\mu_j(t_\star)}{\Delta 
\log t_\star}，
\end{equation}
where $M^c_\star$ is the total mass processed into stars throughout the galaxy history until $t_{\star} =0$, and $\mu_{j}$ is the fraction of  $M^c_\star$.

The time-dependent $specific$ SFR (SSFR) is defined as:
\begin{equation}
\label{eq:SSFR}
       {\rm SSFR}(t_\star) =
       \frac {1}{M^c_\star}
       \frac {d M^c_\star(t_\star)}{dt_\star} \approx
       \frac{\log e}{t_\star}\frac {\mu^c_s(t_\star)}{\Delta \log t_\star}.
\end{equation}
This is a better quantity to describe the SFH for galaxies with different total masses.
The smoothed SSFR in each time box can be constructed as $\mu_{j}$ divided by the length of the time box, which is used to describe the SFHs for ETGs. And in this work, we take time grid from $\log t_\star = 6.0$ to 10.1 in steps of $\Delta \log t_\star = 0.1$ dex.
\begin{figure*}
\begin{center}
 \includegraphics[bb=40 220 780 600,height=15.cm,width=32.cm,clip,angle=0,scale=0.5,angle=0]{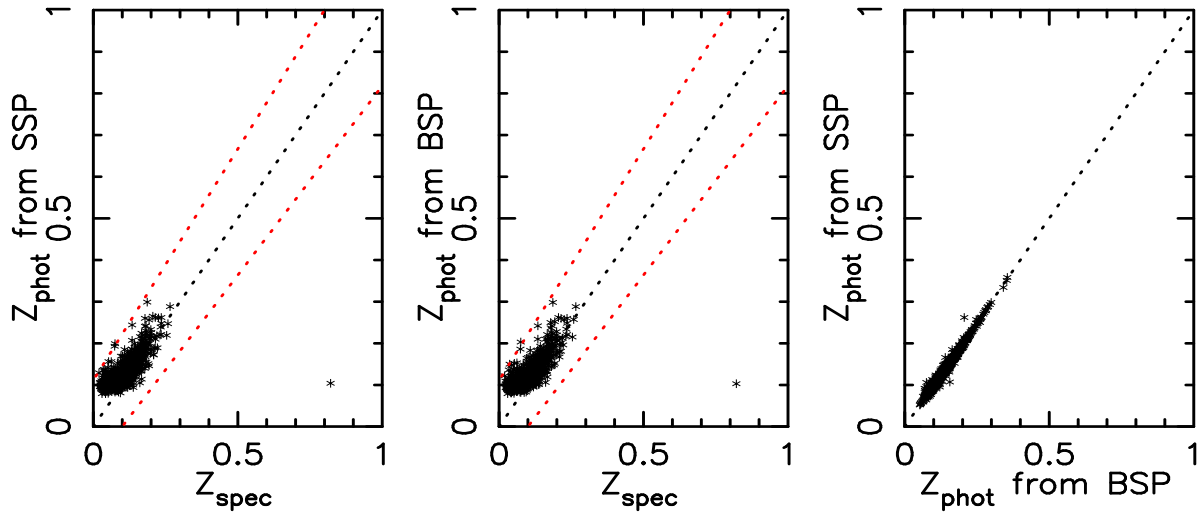}
 \caption{The left and middle panels show the comparisons between  spectroscopic redshift ($z_{spec}$) and photometric redshift ($z_{phot}$) for SSP and BSP models, respectively. The red and black dashed lines are for $z_{phot}-z_{spec}$)$/$($1+z_{spec}$)$<0.1$ and $z_{phot}=z_{spec}$, respectively. The right panel gives the comparison between $z_{phot}$ estimated from BSP and SSP models. The black dashed line is for $z_{phot}$ from BSP $=z_{spec}$ from SSP. }
 \label{fig:photoz_com}
\end{center}
\end{figure*}

\subsection{The EPS models and the theoretical template spectral library}
\subsubsection{The EPS models}

 In order to check the effects of binary interactions on parameter determinations of ETGs, we adopt two EPS models: the single stellar populations \citep[SSPs]{Zhang04} and the binary stellar populations \citep[BSPs]{Zhang05}. These two models present the SEDs of stellar populations with and without binary interactions at 90 ages and 7 metallicities ($Z$ from 0.0001 to 0.03). The ages vary from log($t_{i}\rm /yr$)$=5.000$ to $10.175$.

These two EPS models were built based on the Cambridge stellar evolution tracks \citep{Eggleton71,Eggleton72,Eggleton73}, BaSeL-2.0 stellar atmosphere models \citep{Lejeune97,Lejeune98} and various initial distributions of stars. The Cambridge stellar evolution tracks can be obtained by the rapid single/binary evolution codes \citep{Hurley00,Hurley02}, which is based on the stellar evolutionary track by \citet{Pols98}. In the binary evolution code, various processes are included, such as mass transfer, mass accretion, common-envelope evolution, collisions, supernova kicks, tidal evolution, and all angular momentum loss mechanisms.The main input parameters of the standard models are as follows.\\
(1) The IMF of the primaries gives the relative number of the primaries in the mass range $M \rightarrow M +$ d$M$. The initial primary-mass $M_1$ is given by
\begin{equation}
M_1 = \frac{0.19X}{(1-X)^{0.75} + 0.032 (1-X)^{0.25}},
\label{eq:imfms79-app}
\end{equation}
where $X$ is a random variable distributed uniformly in the range [0, 1].
The distribution is chosen from the approximation to the IMF of \citet{Miller79} as given by \citet{Eggleton89}
\begin{equation}
\phi(M)_{_{\rm MS79}} \varpropto  \left\{ \begin{array}{lll}
                           M^{-1.4}, & 0.10 \leq M \leq 1.00,\\
                           M^{-2.5}, & 1.00 \leq M \leq 10.0,\\
                           M^{-3.3}, & 10.0 \leq M \leq 100.,                 
                 \end{array}
               \right.
\label{eq.imfms79}
\end{equation}

where $M$ is the stellar mass in units of the solar mass M$_{\rm \odot}$.\\
(2) The initial secondary-mass distribution, which is assumed to be correlated with the initial primary-mass distribution, satisfies a uniform distribution
\begin{equation}
n(q)=1.0,\,\,\, 0.0\le q \le 1.0,
\label{eq:nq}
\end{equation}
where $q=M_{2}/M_{1}$.
\\
(3) The distribution of orbital separation (or period) is taken as constant in log$a$ (where $a$ is the separation) for wide binaries and fall off smoothly at close separations
\begin{equation}
a{\rm n}(a) = \left\{ \begin{array}{ll}
                           a{\rm_{sep}} (a/a_{0})^{m}, \, a \leq a_{0}\\
                  a{\rm_{sep}}, \,\,\,\, a_{0} < a <a_{1}
                 \end{array}
               \right.
\label{eq:disa}
\end{equation}
in which $a_{\rm sep} \sim 0.070, a_0 = 10 {\rm R_\odot}, a_1 = 5.75 \times 10^6 {\rm R_\odot}$  $=0.13$pc, and $m \sim 1.2$
\citep[][]{Han95}.\\
(4) The eccentricity distribution satisfies a uniform form $e\,=\,X$, $X\in $[0, 1].\\

Other assumptions and more details are the same as given in \citet{Zhang04} for SSP models and \citet{Zhang05} for BSP models. In these earlier works, the 2.2-version of $BaSeL$ low-resolution library of \citet{Lejeune97,Lejeune98} has been adopted. Meanwhile, in the full-spectrum fitting process, we fit the observed spectra with intermediate resolution. 
We do some modification for these two models, instead of using $BaSeL$ library
we choose the high-resolution ($\sim$\,0.1\,{\AA}) B\tiny{LUERED} \normalsize{library} of \citet[][more details can be found in this paper]{Bertone08} here,whereas other input physics remain unchanged.

\begin{figure*}
\begin{center}
 \includegraphics[bb=15 200 780 600,height=15.cm,width=32.cm,clip,angle=0,scale=0.5,angle=0]{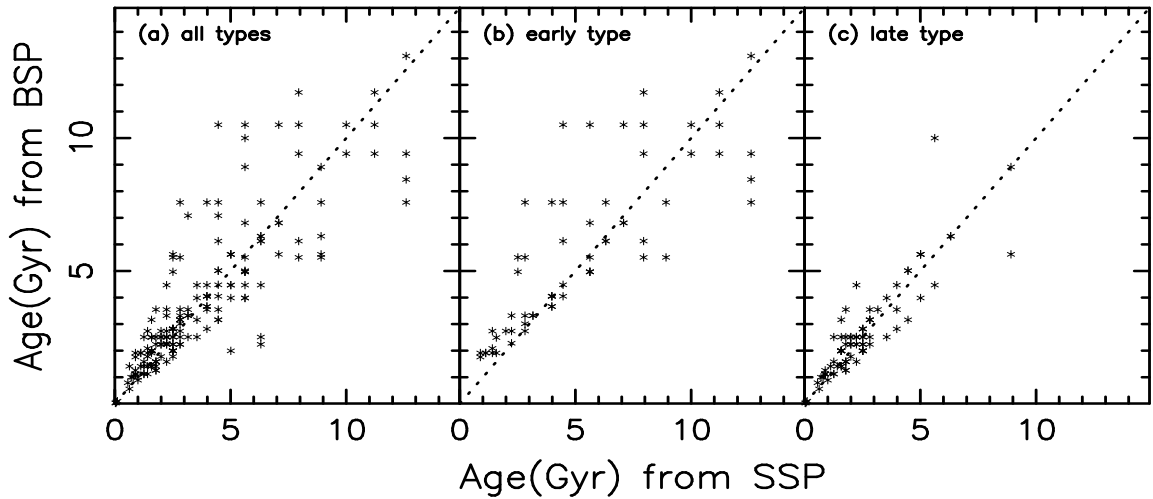}
 
 \caption{Comparisons between the ages obtained from SSP and BSP models for different spectral type galaxies. The left, middle and right panels are for all type, early type, and spiral and irregular galaxies, respectively.  }
 \label{fig:photoage_com}
\end{center}
\end{figure*}

\subsubsection{The teoretical template SED library }
The galaxy template used in the SED fitting method should not only comprise the SEDs of different ages, but also include the SFH of galaxy. The EPS models provide only the SEDs of stellar populations without considering the SFR. Therefore, at a given age we need to generate the SEDs of galaxies with different SFHs by means of EPS models. In this work, our theoretical library include elliptical (E), lenticular (S0), spiral (Sa, to Sd) and irregular (Irr) types.

In several studies, it has been found that the observational properties of local field galaxies with different SFHs can be roughly matched by a population with different SFRs. For example, \citet{Kennicutt86} used an exponentially declining SFR to describe the local spirals and their results could match the observations well. In this work, we built eight types of galaxy: the Burst galaxy with a delta burst SFR, the Irregular with a constant SFR, six types with the following exponentially declined SFR with characteristic time decays, 

\begin{equation}
\label{eq:e-decling-sfr}
\psi(t) \propto exp(-t/\tau) ,
\end{equation}
where $\tau$ and $t$ are the $e-$folding time scale and the age of population, respectively. In which, $\tau =1, 2, 3, 5, 15,$ and $30$ Gyr are for E, S0, Sa to Sd, respectively. \citet[][BC03]{Bruzual03} software package provides the code to construct the galaxy template with different SFRs. With the stellar population models of SSP and BSP and the $e-$folding SFR, the BC03 package has been used to build the galaxy templates. $Z=0.02$ has also been assumed, and the age range is from 10$^{5}$ yr to 10$^{10.175}$ yr.

\section{EFFECTS OF BINARY INTERACTIONS ON PHOTO-z DETERMINATIONS} 
Binary interactions could change the $UV$ flux significantly. Based on the SSP and BSP population models, by analyzing the photometric data with the HyperZ, the photo-z, the spectral type (corresponding to the galaxy morphological type) and the age in Gyr can be obtained. Below we will discuss the effects of binary interactions on these parameter determinations.

For checking the accuracy of the photometric redshift $z_{phot}$, we compare $z_{phot}$ with the spectroscopic $z_{spec}$ obtained from SDSS DR7. Comparisons between SDSS $z_{spec}$ and $z_{phot}$ for the SSP and BSP model are shown in left and middle panels of Fig.\ref{fig:photoz_com}, respectively. It can be seen that our $z_{phot}$ obtained with the SSP and BSP models are in good agreement with $z_{spec}$. All galaxies satisfy the relation ($z_{phot}-z_{spec}$)$/$($1+z_{spec}$)$<0.1$. 

We also compare $z_{phot}$ estimates of galaxies between the SSP and BSP models in the right panel of Fig.\ref{fig:photoz_com}. Here the $z_{phot}$ estimates are approximately the same for both models. This indicates that binary interactions have negligible effect on the photo-$z$ estimates. Since the young stellar populations can also radiate the UV-light, which imitate the effect of binary interactions.
The changes induced on the SEDs of populations by the binary interactions are compensated in most cases by the young stellar populations, and thus obtain the same photo-$z$ for the SSP and BSP models.

As described above, using the HyperZ code we can also obtain the spectral types and ages of galaxies. In this Section, we analyse the influence of binary interactions on the age estimates for different spectral type galaxies. The comparisons for age obtained from the SSP and BSP models for eight type galaxies, early type galaxies, and spiral and irregular galaxies are displayed in the left, middle and right panels in Fig. \ref{fig:photoage_com}, respectively. It can be found that the ages of early type galaxies are large, and those of the spiral and irregular galaxies are small. This is consistence with that the early type galaxies could be almost composed of old stars, while the spiral and irregular galaxies also have star formation activities. From Fig. \ref{fig:photoage_com}, it can be seen that the age estimated by the BSP models are larger than that determined by the SSP models for early galaxies, and there is no difference for the case of spiral and irregular galaxies. It is because that the existence of younger stellar populations in spiral and irregular galaxies also have the same effect on UV spectra, which pollute the effect of binary interactions. Once binary interactions are neglected in the EPS models, the age of early type galaxies will be underestimated, by contrast, the effects on the age estimation can be negligible for other type galaxies.

\section{EFFECTS OF BINARY INTERACTIONS ON PARAMETER ESTIMATES FOR ETGS} 
STARLIGHT is for fitting the full spectrum with the EPS models to derive physical properties of stellar population systems. For checking the reliability and robustness of STARLIGHT, \citet{CidFernandes05} have analysed the spectra of mock galaxies whose stellar populations were already known, their results have good agreement with the known components \citep[for details see ][]{CidFernandes05}.
By using the STARLIGHT code to fit the SDSS spectra of ETG sample selected from \citet{Fukugita07}, one can obtain (i) the average stellar population age, (ii) the average stellar metallicity, (iii) the fraction of mass contribution for each stellar population, (iv) the velocity dispersion and (v) the extinction. To investigate the effects of binary interactions, two EPS models (SSP and BSP models) have been used to fit the observed spectra of ETGs.

\subsection{The effects on estimates of average age and metallicity}

\begin{figure*}
\begin{center}
 \includegraphics[bb=30 20 700 550,height=13.cm,width=15.cm,clip,angle=0,scale=0.5,angle=0]{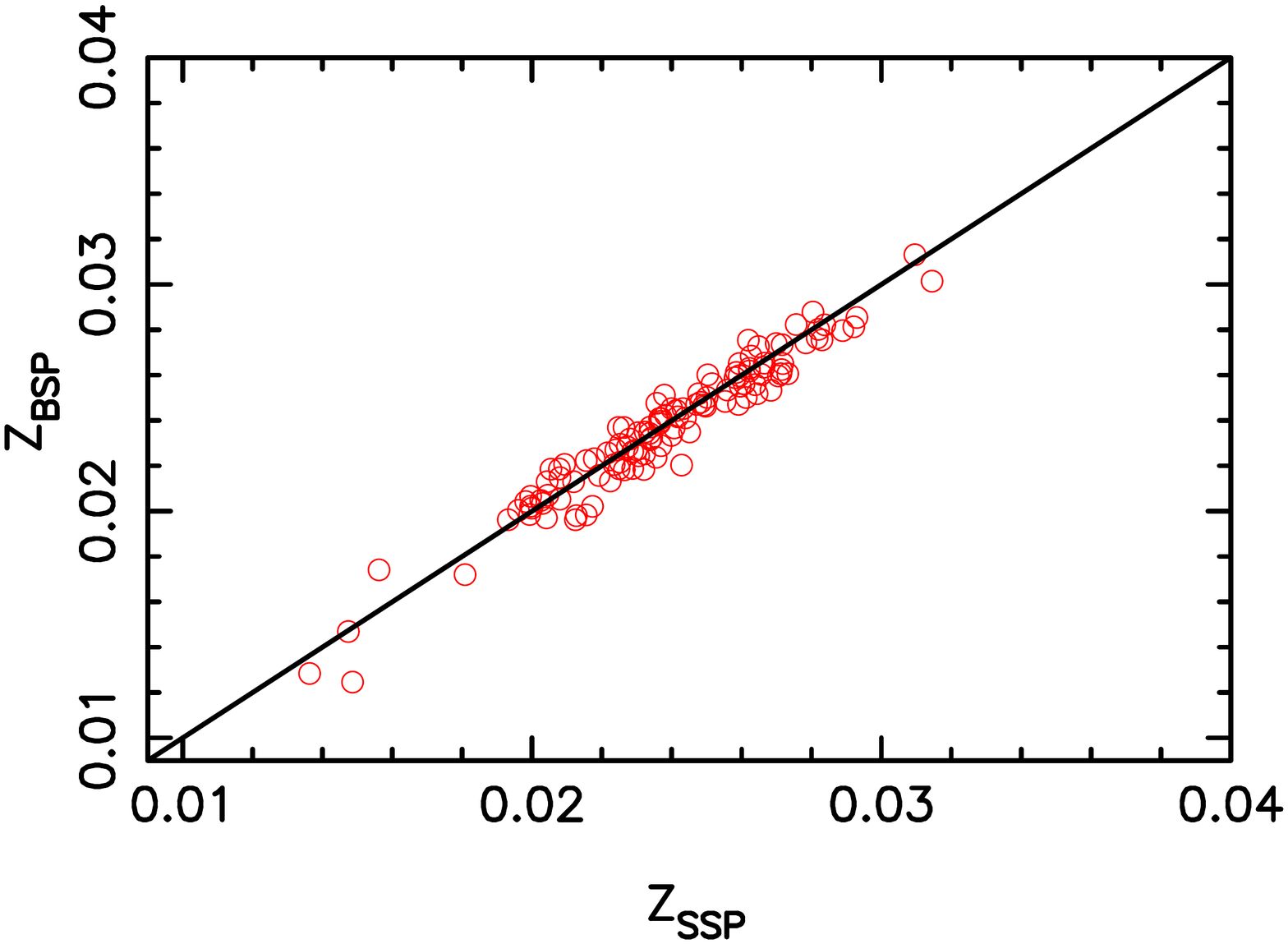}
 \includegraphics[bb=30 20 700 550,height=13.cm,width=15.cm,clip,angle=0,scale=0.5,angle=0]{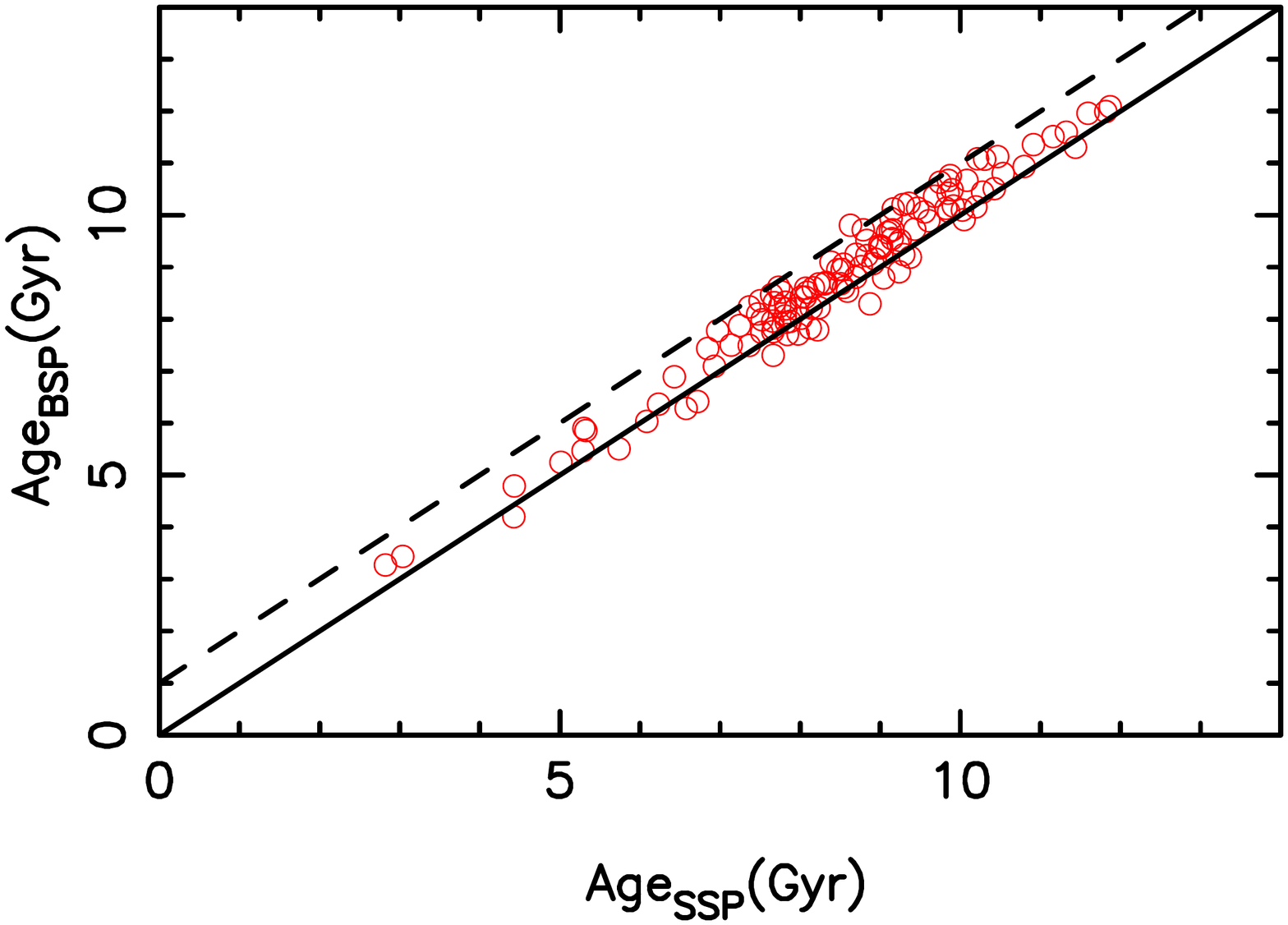}
 \caption{Comparisons of the SSP- and BSP-fitted paramters for the ETG sample. The left and right panels present the comparison for mean metallicities and ages, respectively. The red dots illustrate the ETGs, the dashed lines in right panle are for $\vert$Z$\rm_{SSP}-$Z$\rm_{BSP}\vert=0.001$, the dashed line in left panel is for Age$\rm _{BSP}-$Age$\rm _{SSP}=1.0$ Gyr.}
 \label{fig:age_com}
\end{center}
\end{figure*}

\begin{figure*}
\begin{center}
 \includegraphics[bb=17 100 800 600,height=15.cm,width=30.cm,clip,angle=0,scale=0.5,angle=0]{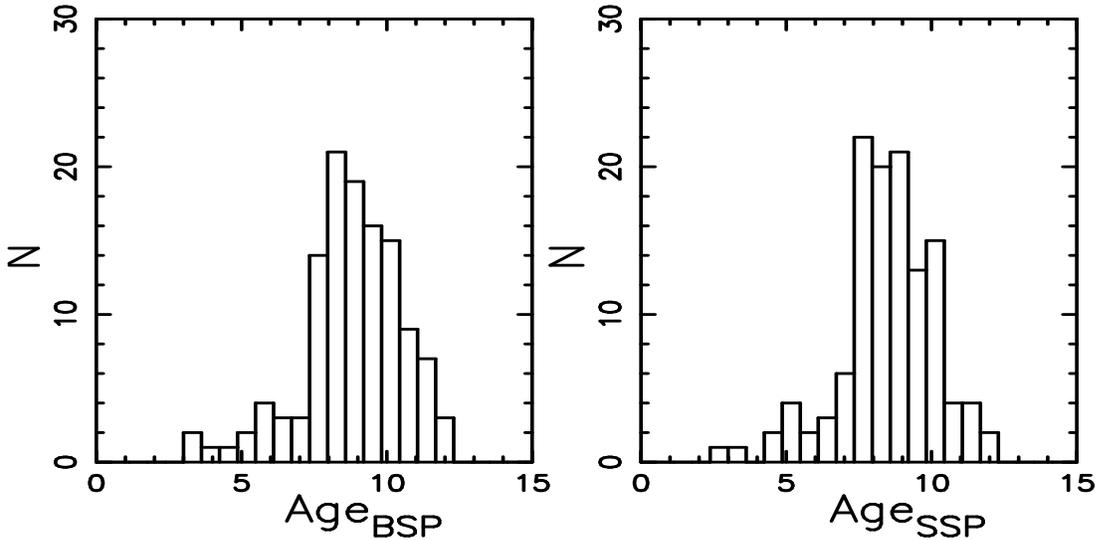}
 
 \caption{The age distributions for all ETGs sample. The left and right panels are present the result for BSP and SSP models, respectively.}
 \label{fig:age_hist}
\end{center}
\end{figure*}

The mean age and metallicity can be used to characterize the stellar population mixture of a galaxy. STARLIGHT presents both the light- and mass-weighted mean metallicity and age. In this section, we adopt the first light-weighted mean age, for it has direct relation with the observed spectrum. We show the clear comparisons for age and metallicity estimates between two models in Fig. \ref{fig:age_com}.
 
Fig. \ref{fig:age_com} shows the comparisons of the SSP- and BSP-fitted parameters for the ETG sample. The left and right panels are for the mean metallicity and age, respectively. It can be seen that the metallicity estimated by the SSP and BSP models are approximately equal: $\vert$Z$\rm_{SSP}-$Z$\rm_{BSP}\vert<0.001$ for almost all of ETGs. These indicate that the binary interactions have negligible effect on the metallicity estimates.

On the other hand, it has been found that the BSP model gives older mean ages for ETGs obviously than the SSP model. Our results show that the BSP-fitted ages in $\sim33.3\%$ of 120 ETG sample are around $0.5-1.0$\,Gyr larger than the SSP-fitted ages; $\sim44.2\%$ are only $0.1-0.5$\,Gyr larger; the rest $\sim22.5\%$ are approximately equal. In Fig. \ref{fig:age_hist}, we also present the age distribution estimated by the SSP and BSP models. The left and right panels are for the BSP and SSP models, respectively. From these distributions, one can see that the age estimates are centered at about 9.0 Gyr and 8.0 Gyr for the BSP and SSP models, respectively. The mean ages for the ETG sample are about 8.9 and 8.4 Gyr for the BSP and SSP models, respectively.
Thus, the binary interactions have much effect on the age estimate. Once we neglect binary interactions in the EPS models, the age of ETGs will be underestimated.

\subsection{The effects on estimates of SFH}
SFH can be characterized in terms of fundamental properties of galaxies, which is described by the SFR evolved with time. In this work, we use SSFR evolving with time to characterize SFH of ETG sample. As described in Sec. 3.1.2, the smoothed method has been adopted (eqs. \ref{eq:SFR} and \ref{eq:SSFR}) to describe the SFH. The smoothed SSFR in each time box is computed as the mass-weighted vector of the corresponding stellar population divided by the length of the time box. 

\begin{figure*}
\begin{center}
 \includegraphics[bb=18 38 780 600,height=25.cm,width=35.cm,clip,angle=0,scale=0.5,angle=0]{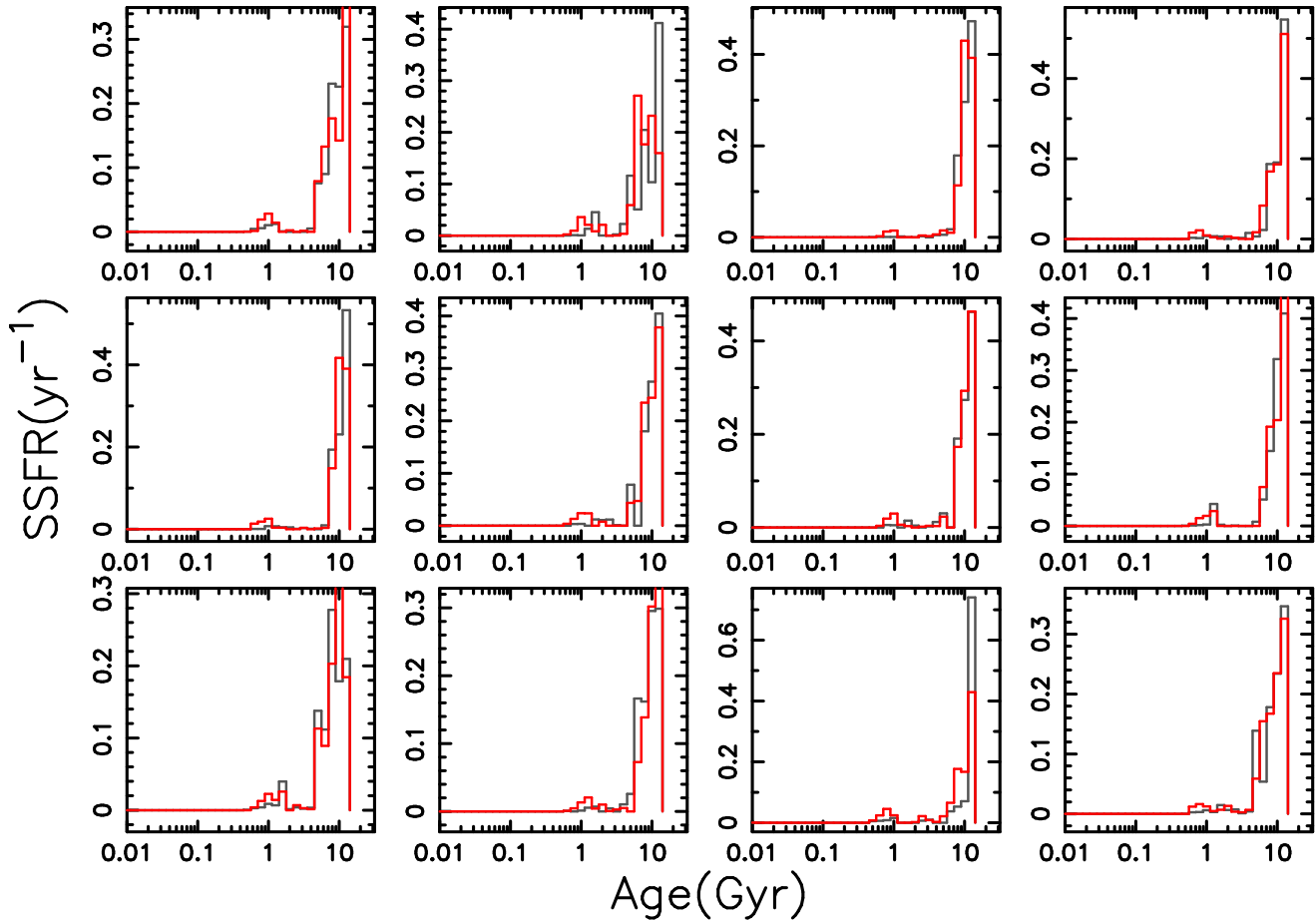}
 \caption{The SSFR fitted with the SSP and BSP models for the A sub-sample. The red and grey lines represent for SSP and BSP models, respectively. }
 \label{fig:sfr_com1}
\end{center}
\end{figure*}

\begin{figure*}
\begin{center}
 \includegraphics[bb=18 20 780 600,height=28.cm,width=35.cm,clip,angle=0,scale=0.5,angle=0]{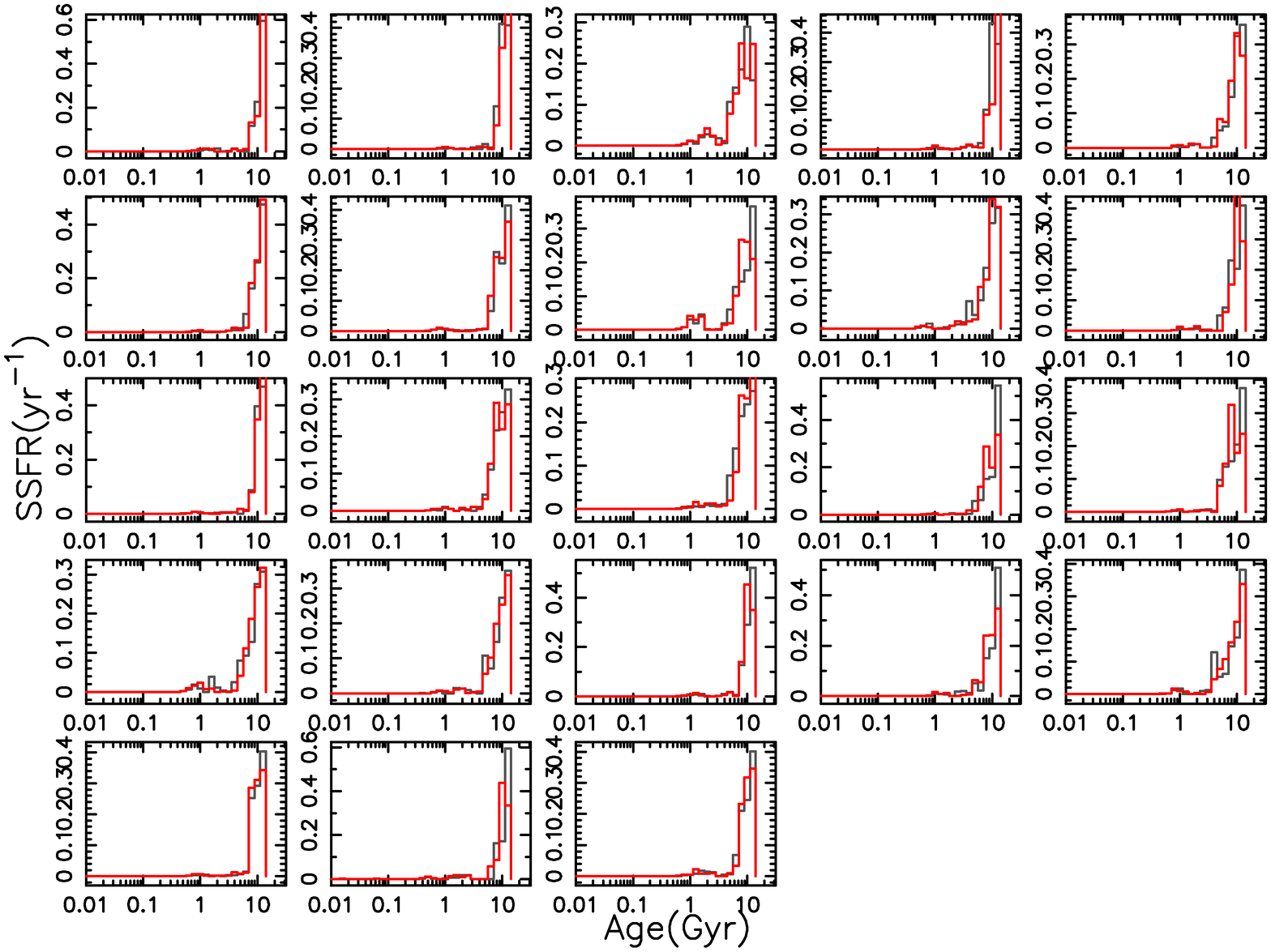}
 \caption{The SSFR fitted with the SSP and BSP models for the B sub-sample. The red and grey lines represent for SSP and BSP models, respectively. }
 \label{fig:sfr_com2}
\end{center}
\end{figure*}

\begin{figure*}
\begin{center}
 \includegraphics[bb=18 60 780 590,height=23.cm,width=35.cm,clip,angle=0,scale=0.5,angle=0]{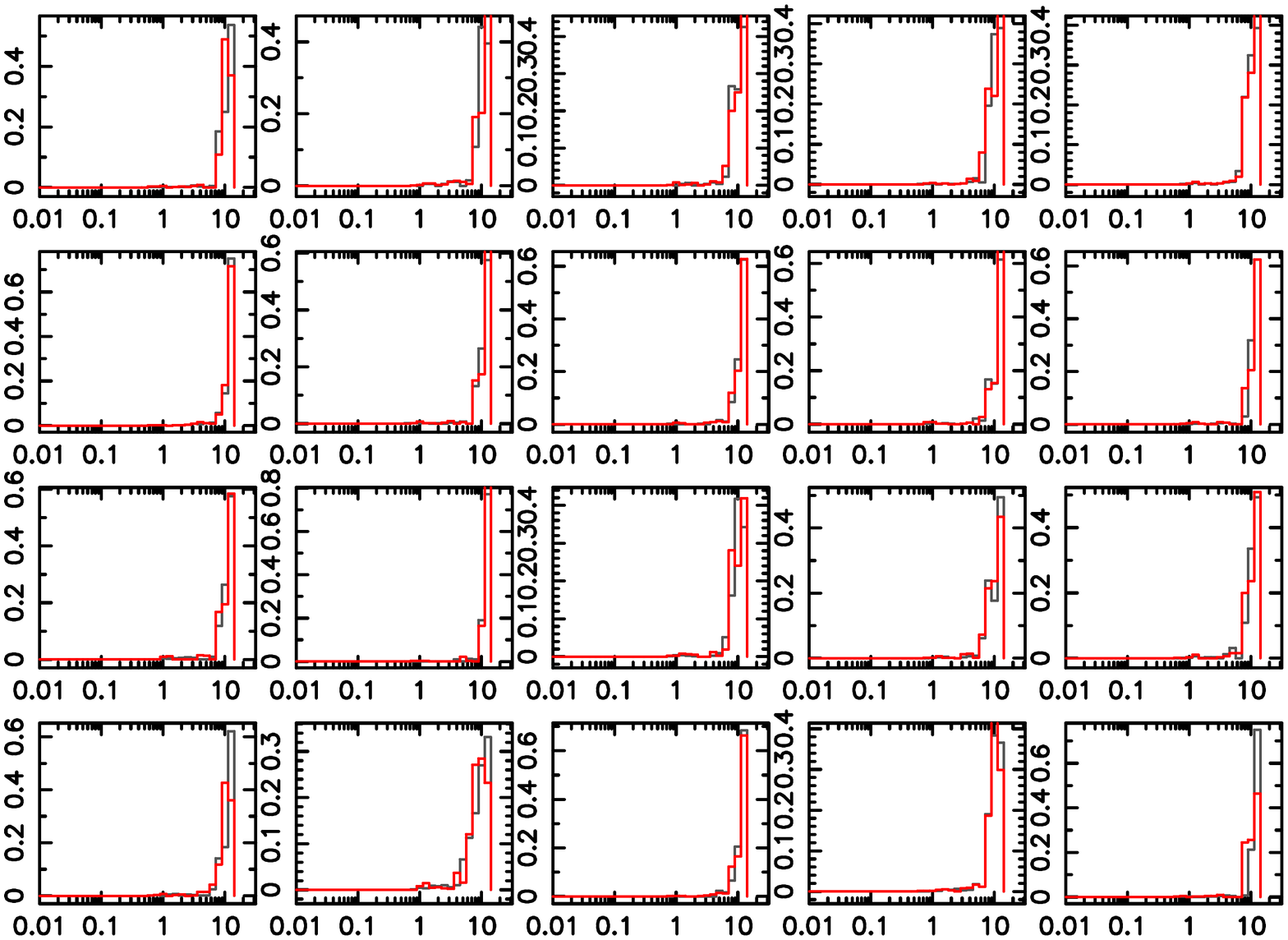}
\includegraphics[bb=18 31 780 590,height=23.cm,width=35.cm,clip,angle=0,scale=0.5,angle=0]{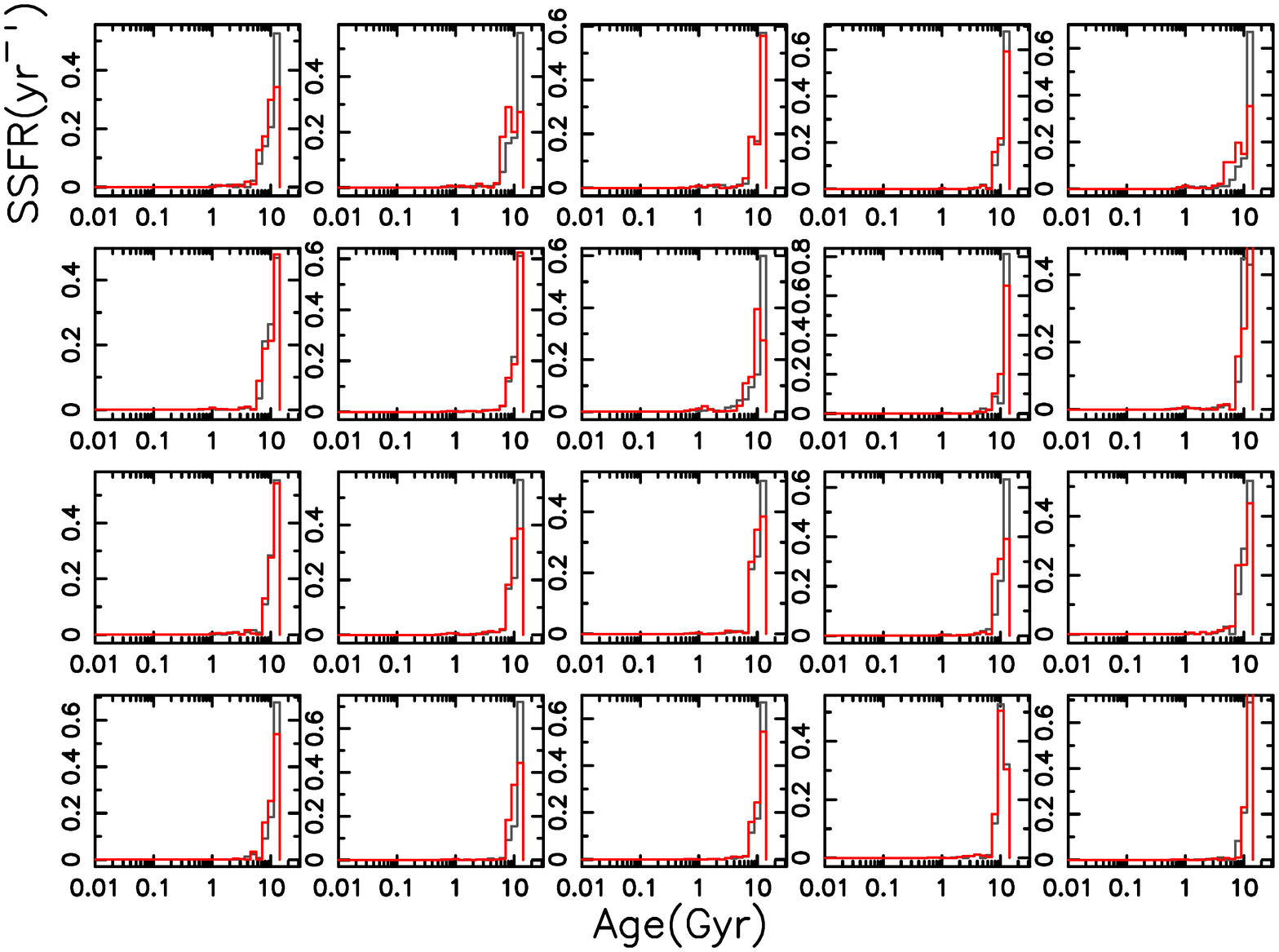}

 \caption{The SSFR fitted with the SSP and BSP models for the C sub-sample. The red and grey lines indicate for SSP and BSP models, respectively. }
 \label{fig:uv_weak}
\end{center}
\end{figure*}

In order to investigate the effects of binary interactions on the SSFR estimates, we divide the sample into A, B and C three sub-sample. Fig. \ref{fig:sfr_com1}, \ref{fig:sfr_com2} and \ref{fig:uv_weak} show the SSFR for each galaxy as a function of evolutionary time for A, B and C sub-sample, respectively. For each galaxy of the ETG sample, SSFR estimated from the full-spectrum fitting with the SSP model is indicated by a red solid line, while SSFR estimated with the BSP model by a gray solid line, respectively. These figures give direct comparisons of SFH of the ETG sample reconstructed with the SSP and BSP models. These allow us to investigate the effects of binary interactions on the SFH estimation for galaxy.
   
From these figures, it can be seen that the tendency of SSFRs estimated with the SSP and BSP models for each galaxy of the ETG sample has a good agreement for all ETGs, relative low SSFR in the late evolution stage and high SSFR in the early epoch. For all ETGs, the early period is very important for their star formation. Almost all ETGs have already assembled the bulk of their stellar masses over $5$ Gyr ago. This is consistence with that the ETGs are thought to be almost exclusively composed of old stars. 
The difference of SSFR between these two models happens in the late evolution stage with age$\leq 1.0$\,Gyr. We divide the ETG sample into three sub-sample to analyse this difference. 

\begin{itemize}
\item
For the A sub-sample (12 ETGs, $\sim10.0\%$ of ETGs), we show the comparison of SSFRs between the SSP and BSP models in Fig. \ref{fig:sfr_com1}. For the late evolution stage, all ETGs in the A sub-sample have low SSFRs for the SSP model and they may have some star formation in this stage.
SSFRs estimated from the BSP model for ETGs are almost close to zero in the late evolution epoch, and ETGs have no star formation in this epoch. From this comparison, it can be found that there is a large difference for the SSFR at late evolution epoch between the SSP and BSP models for the A sub-sample.
\item
For the B sub-sample (23 ETGs, $\sim19.2\%$), we present the SSFRs estimated from both SSP and BSP models in Fig. \ref{fig:sfr_com2}. All ETGs have very low SSFRs in the late evolution stage for both SSP and BSP models. 
For these ETGs fitted with the BSP and SSP models, they may have some small star formation at late evolution epoch. 
\item
For the C sub-sample (85 ETGs, $\sim70.8\%$), we give SSFRs obtained from both SSP and BSP models in Fig.  \ref{fig:uv_weak}. SSFRs are all almost close to zero for ETGs fitted with two models. For these ETGs, there is no star formation in the late evolution epoch for both SSP and BSP models and it has no difference for the SSFR at this epoch between two models.
\end{itemize}

From the above anylases, we find that there exists difference for SSFR estimates between the SSP and BSP models for the A sub-sample, and it has no difference between the SSP and BSP models for the B and C sub-samples. For the A sub-sample, some star formations exist in the late evolution stage for the SSP model, while there is no star formation in the late evolution stage for ETGs fitted with the BSP model. The reason is that the binary interactions can enhance the UV flux, which is similar to the effect of young stars. Thus the inclusion of binary interactions in the EPS models lead to a lower SSFR estimate. This also indicate that ETGs in the A sub-sample may have some UV emission, and such UV emission can be explained by young stars in the SSP model and through binary interactions in the BSP model. For the B sub-sample, ETGs have small star formation in late evolution stage estimated from both SSP and BSP models. For these ETGs, they may have residual star formation. For the C sub-sample, all ETGs have no star formation in late evolution epoch fitted by both SSP and BSP models. ETGs in the C sub-sample may have neither UV emission nor star formation in the late evolution stage.

\begin{figure*}
\begin{center}
 \includegraphics[bb=18 60 780 590,height=23.cm,width=35.cm,clip,angle=0,scale=0.5,angle=0]{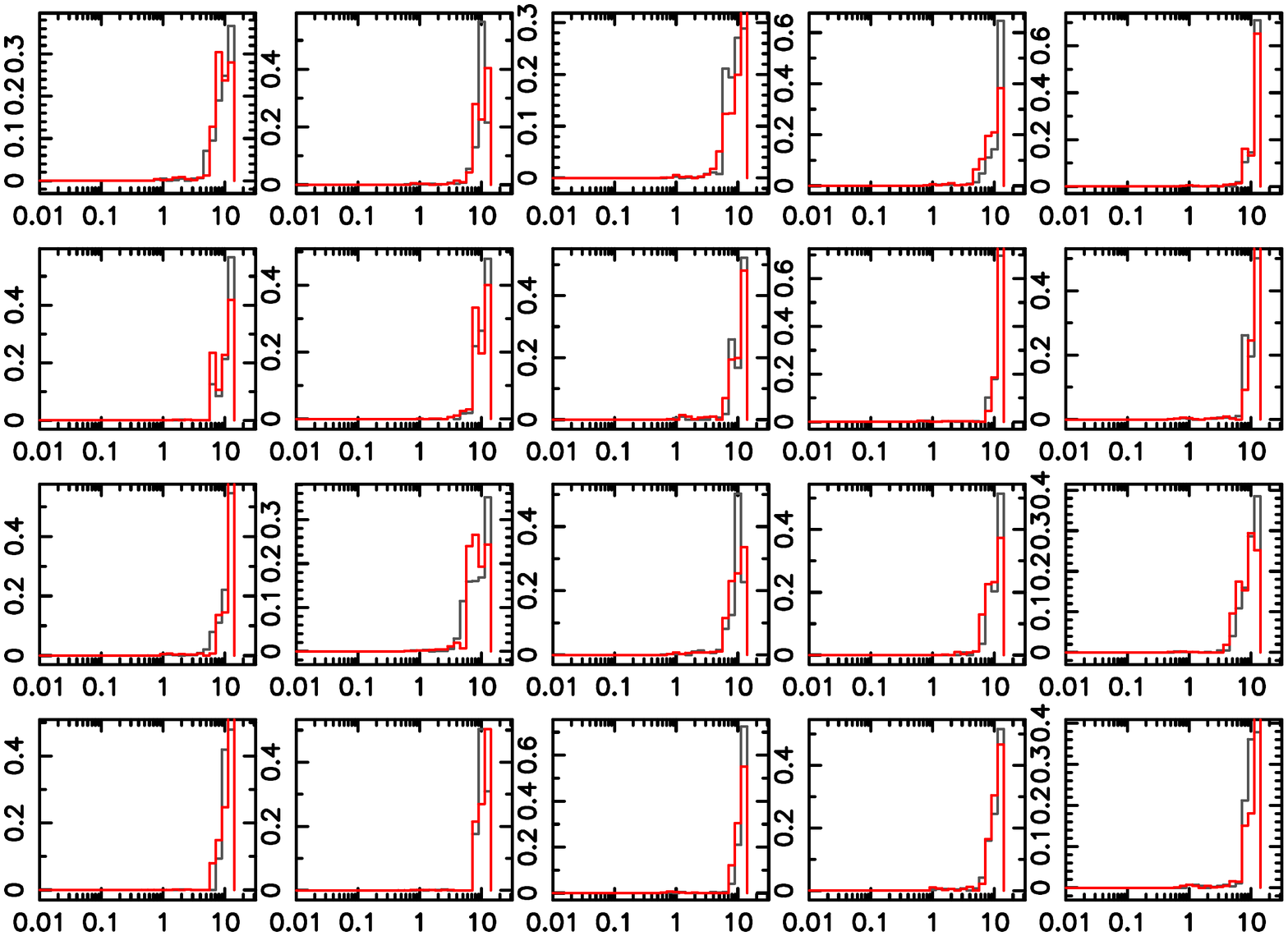}
\includegraphics[bb=18 28 780 590,height=23.cm,width=35.cm,clip,angle=0,scale=0.5,angle=0]{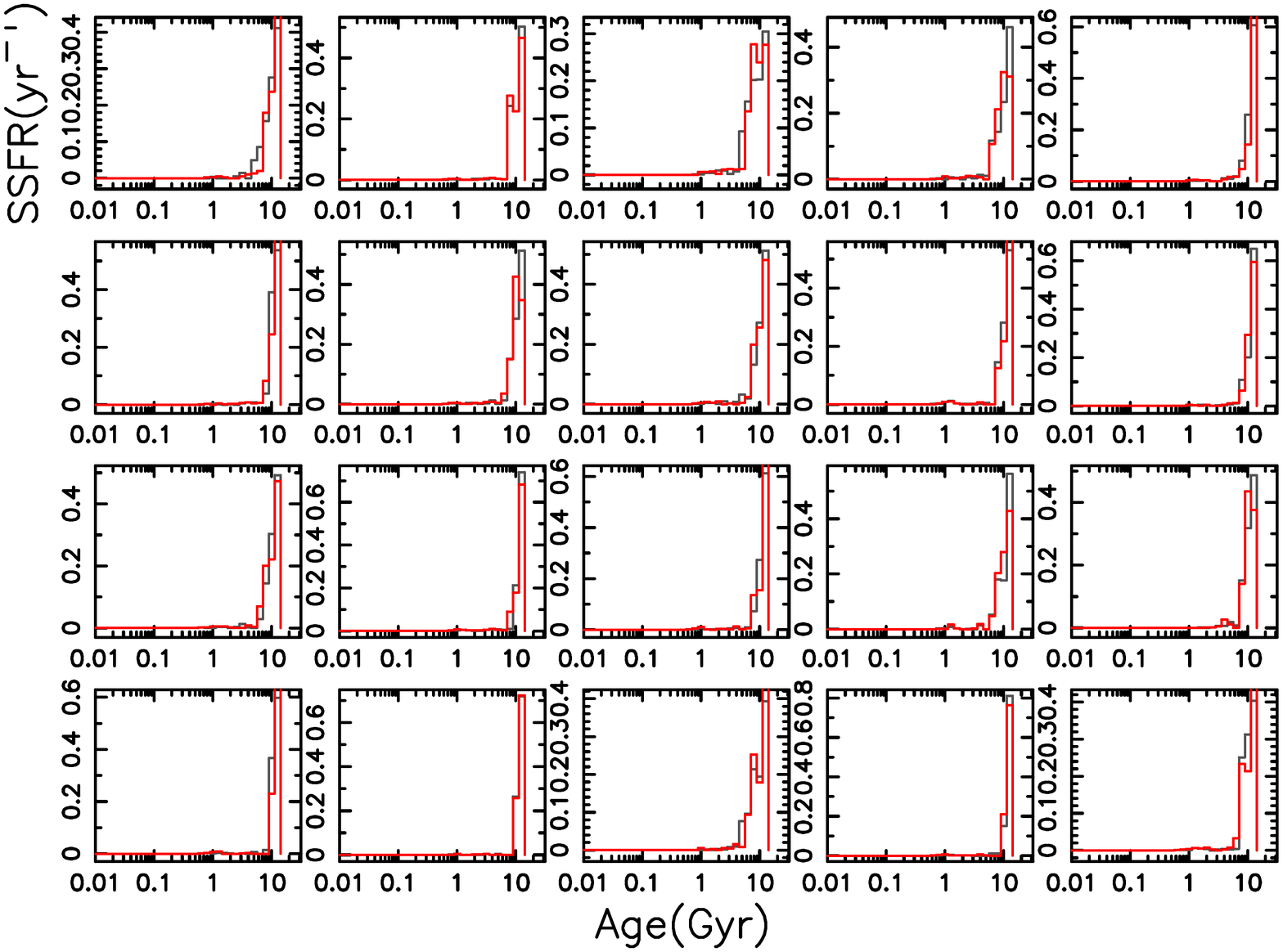}
\textbf{Figure 7.} Continue 

\end{center}
\end{figure*}

\begin{figure*}
\begin{center}
 \includegraphics[bb=10 400 780 590,height=8.cm,width=35.cm,clip,angle=0,scale=0.5,angle=0]{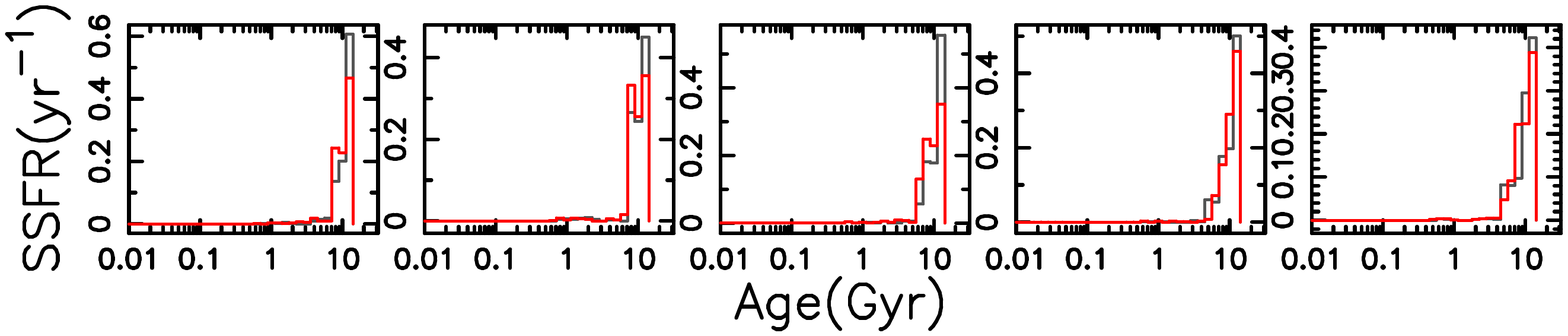}
\textbf{Figure 7.} Continue 
 
\end{center}
\end{figure*}

\subsection{Dependence on the EPS models}
The reliability of the results may depend on the EPS models. The main ingredients of the EPS models include the stellar evolution track, the stellar spectral library and the IMF form. \citet{zfh12} chose the different IMF form of \citet{Miller1979} and \citet{Salpeter1955} to analyse the influence of IMF on the SFR estimation in terms of UV luminosity for different type of galaxies, and they found that the effect on the UV luminosity caused by the variation in the IMF forms was smaller than that between the SSP and BSP models for ETGs. \citet{Hernandez14} also used the IMF of  \citet{Salpeter1955} and \citet{Chabrier2003} to compare UV colours of sample galaxies. They found that, although the \citet{Chabrier2003} IMF model was slightly bluer in Nuv$-r$ than the \citet{Salpeter1955} IMF model, both models with different IMF could match with the observed colours of ETGs. All these certify that the IMF choose has small effect on our above results.

In our previous work \citep{Zhang12}, we used three EPS models to investigate the dependence of parameter determinations on the adoption of EPS model, and found that the parameter determinations were independence of the EPS models. All this indicates that our above results are independent of the main ingredients of EPS models.   
\section{SUMMARY AND CONCLUSIONS}
Applying the HyperZ SED fitting algorithm and the template SED library built with the SSP and BSP models, we present the photo-$z$, age and spectral type estimates for the photometric data sample, which include $4278$ galaxies with all $F_{UV}$, $N_{UV}$ and $ugriz$ magnitudes and are randomly selected from SDSS DR7.
We compare the SDSS $z_{spec}$ and $z_{phot}$ obtained from the SSP and BSP models in order to check the accuracy of photometric redshift. From these comparison, we find that the $z_{phot}$ are in good agreement with the $z_{spec}$ for SSP and BSP models, and all galaxies have ($z_{phot}-z_{spec}$)$/$($1+z_{spec}$)$<0.1$. 
From the comparison between SSP and BSP models for the $z_{phot}$ estimations of galaxies, we find that the binary interactions have negligible effect on the photo-$z$ estimations for galaxy sample.
We also compare the age estimates between the SSP and BSP models for different spectral type galaxies. From these comparisons, we find that the effect of binary interaction on the age estimation is obviously for early type galaxies, and has no effect for other type galaxies. Once binary interactions are neglected in the EPS models, the age of early type galaxies will be underestimated, by contrast, the effects on the age estimation can be negligible for other type galaxies. 

In order to give a detailed analysis for the effects of binary interactions on the property determinations for ETGs, 
based on the STARLIGHT code of \citet{CidFernandes05}, we present a detailed stellar population synthesis for the ETG sample selected from the \citet{Fukugita07} catalogue. The spectra of these ETGs are obtained from the SDSS DR7 data and have Petrosian magnitudes in the $r$ band brighter than 16 mag.

We fit the spectra of ETG sample with the SSP and BSP models, respectively. From these spectra fitting, we obtain the average age and metallicity from the SSP and BSP models for each ETG of the sample. Our results show that the mean metallicity from the two different models are approximately equal. The comparisons for mean age show that the BSP-fitted ages in $\sim33.3\%$ of ETG sample are around $0.5-1.0$\,Gyr larger than the SSP-fitted ages; $\sim44.2\%$ are only $0.1-0.5$\,Gyr larger; the rest $\sim22.5\%$ are approximately equal.
Our results suggest that binary interactions can possibly play an important role for age estimation for ETGs. It is necessary to take binary interactions into account in the EPS models.

We use the smoothed method to derive the smooted SSFR, and then reconstruct the SFH for each galaxy of the ETG sample by the smoothed SSFR. For all ETGs, we find that the early evolution stage is a very important period for star formation for both SSP and BSP models, most of the stars are formed at the early evolution epoch. From the comparison of SSFR estimates between the SSP and BSP models, we find that the difference of the SSFR between the SSP and BSP models is large at the late evolution stage for ETGs. 
In this paper, we divide the ETG sample into A, B and C three sub-sample. The results show that there exists  difference for SSFR estimates between the SSP and BSP models for the A sub-sample, and it has no difference for the B and C sub-samples. We conclude in the following. (i) For the A subsample ($12$ ETGs, $\sim10.0\%$ of ETGs), ETGs may have UV emission. Some star formations exist in the late evolution stage for ETGs fitted with the SSP model, they has no star formation in the late evolution stage fitted with the BSP model. The reason is that the binary interactions can enhance the UV flux. (ii) For the B sub-sample ($23$ ETGs,  $\sim19.2\%$), ETGs may have some small star formations at the late evolution stage. The reason is that all ETGs have very low SSFRs in the late evolution epoch for both SSP and BSP models. (iii) For the C sub-sample ($85$ ETGs, $\sim70.8\%$), ETGs may have neither star formation nor UV emission in the late evolution stage. Therefore the SSFRs are all close to zero in the late evolution estimated from both SSP and BSP models.

\section*{Acknowledgments}
We appreciate the anonymous referee for reviewing our paper very carefully and relevant comments on our work.
This work is supported by the program of the Light in China' Western Region (LCWR, Grant No. XBBS201221) and the National Natural Science Foundation of China (Grant Nos. 11303080, 11103054 and 11273053).

\bibliography{ms}

\label{lastpage}
\end{document}